# Enhanced third-harmonic generation empowered by doubly degenerate quasi-bound states in the continuum

Tingting Liu,[1,2] Meibao Qin,[3,4] Jumin Qiu,[4,*] Xu Tu,[2] Huifu Qiu,[2]
Feng Wu,[5] Tianbao Yu,[4] Qiegen Liu,[1] and Shuyuan Xiao[1,2,†]

[1]*School of Information Engineering,*

*Nanchang University, Nanchang 330031, China*

[2]*Institute for Advanced Study, Nanchang University, Nanchang 330031, China*

[3]*School of Education, Nanchang Institute of*

*Science and Technology, Nanchang 330108, China*

[4]*School of Physics and Materials Science,*

*Nanchang University, Nanchang 330031, China*

[5]*School of Optoelectronic Engineering,*

*Guangdong Polytechnic Normal University, Guangzhou 510665, China*

## Abstract

Recent advancements in nonlinear nanophotonics are driven by the exploration of sharp resonances within high-index dielectric metasurfaces. In this work, we leverage doubly degenerate quasi-bound states in the continuum (quasi-BICs) to demonstrate robust enhancement of third-harmonic generation (THG) in silicon metasurfaces. These quasi-BICs are governed by $C_{4v}$ symmetry and therefore can be equally excited with the pump light regardless of polarization. By tailoring the geometric parameters, we effectively control $Q$-factors and field confinement of quasi-BICs, and thus regulate their resonantly enhanced THG process. A maximum THG conversion efficiency up to $1.03 \times 10^{-5}$ is recorded under a pump intensity of 5.85 GW/cm². Polarization-independent THG profile is further confirmed by mapping its signal across the polarization directions. This work establishes foundational strategies for the ultracompact design of robust and high-efficiency photon upconversion systems.

\* qiujumin@email.ncu.edu.cn

† syxiao@ncu.edu.cn

## I. INTRODUCTION

Optical metasurfaces, composed of two-dimensional arrays of subwavelength structures, have emerged as a transformative technology for realizing ultrathin planar devices. They offer unprecedented capabilities for precise and flexible control over scattering amplitude, phase, and polarization, thereby driving significant progress in integrated optics, sensing, and information processing[1, 2]. A particularly notable feature of metasurfaces is their ability to enhance and localize electromagnetic fields at the nanoscale, which opens new possibilities for extending the freedom of frequency via nonlinear optical processes[3, 4]. Nonlinear metasurfaces not only alleviate the phase-matching requirements typically imposed in bulk optical media, but also exhibit efficient frequency conversion and flexible control over nonlinear optical responses, which facilitates unique functionalities for applications such as nanoscale light sources[5, 6], nonlinear holography[7–9], ultrafast optical switching[10], and quantum state engineering[11, 12].

A major focus in the field has been boosting nonlinear conversion efficiency by enhancing light-matter interactions and confining electromagnetic fields within the metasurface. Recent progress has shifted towards dielectric metasurfaces, which feature high refractive indices, low losses, significant nonlinear susceptibilities, and high damage thresholds[13]. These metasurfaces support optically induced resonances, enabling novel mechanisms for subwavelength control of nonlinear light-matter interactions[14–16]. In a broad context, dielectric metasurfaces, that support magnetic mode[17, 18], toroidal mode[19, 20], anapole mode[21–23], as well as Fano resonance[24, 25] and electromagnetically induced transparency[26], have been extensively studied for enhanced field confinement in small volumes, yielding pronounced nonlinear optical effects. Bound states in the continuum (BICs), representing localized electromagnetic states embedded within the radiation continuum, further provide a power physical mechanism for confining electromagnetic energy in open resonators[27, 28]. Breaking in-plane symmetry in dielectric metasurfaces enables the transformation of BICs into quasi-BICs, which can be excited by external fields and exhibit sharp resonances with extraordinarily high $Q$-factors, resulting in significantly enhanced field confinement. The $Q$-factors of quasi-BICs depend on the geometric asymmetry of metasurfaces, which offers design flexibility for controlling over nonlinear light-matter interactions[29–31]. Building on these capabilities, quasi-BICs have been extensively exploited in a range of nonlinear optical phenomena, including harmonic generation[32–38], sum-frequency generation[39–41], four-wave mixing[42–44], and entangled photon-pair generation[45–48]. Despite these advantages, quasi-BICs often rely upon symmetry-broken metasurfaces, making them sensitive to pump polarization, which poses challenges to their robustness and limits their practical applicability in polarization-independent systems.

In this work, we report the polarization-independent enhancement of third-harmonic generation (THG) in silicon metasurfaces enabled by doubly degenerate quasi-BICs. By introducing perturbations that break translation symmetry while preserving $C_{4v}$ rotational symmetry, quasi-BICs are generated through the disruption of the ideal conditions of genuine BICs, while fully retaining the polarization-independent degenerate behavior in both linear and resonantly enhanced nonlinear responses. The high $Q$-factor and enhanced field confinement facilitate the efficient nonlinear frequency conversion from the near-infrared to visible light, achieving a maximum conversion efficiency up to $1.03 \times 10^{-5}$ under a pump intensity of 5.85 GW/cm². Finally, we demonstrate the polarization-independent profile of THG power, showcasing the exceptional robustness of the doubly degenerate quasi-BIC metasurfaces in nonlinear optical processes.

## II. RESULTS AND DISCUSSION

The concept of the polarization-independent quasi-BIC metasurfaces for THG enhancement is schematically illustrated in Fig. 1a. The metasurfaces consist of an array of silicon nanodisks arranged in a square lattice that exhibit a $C_{4v}$ rotational symmetry. The unit cell adopts a diatomic design, characterized by a periodicity of $P$ = 990 nm, a height of $H$ = 220 nm, and nanodisk radii denoted as $R_1$ and $R_2$. These geometric parameters are optimized to tune the quasi-BICs to wavelengths in the near-infrared range, thereby avoiding pump absorption in the silicon material[49]. The design is firstly numerically implemented using the finite element method (FEM), referring to the Supporting Information for detailed description of COMSOL numerical simulations. For the case where $R_1 = R_2 = 250$ nm, the $C_{4v}$ symmetry of the metasurface results in the emergence of a pair of doubly degenerate BICs with a frequency of 194.13 THz marked by a red point at the Γ point. These modes can be decomposed into two orthogonal dipole-like modes, which are 90° rotated versions of each other, as shown in Fig. 1b. When a perturbation is introduced such that $\Delta R = R_1 - R_2 = 50$ nm, the BICs will degrade into quasi-BICs with a frequency of 200.79 THz, also marked by a red point at the Γ point. This perturbation breaks the original translation symmetry while still preserving the $C_{4v}$ symmetry of the metasurfaces. Consequently, the degenerate quasi-BICs retain electric field distributions similar to those of original degenerate BICs, as shown in Fig. 1c. The preserved symmetry ensures polarization-independent degenerate behavior in both linear and nonlinear optical responses.

Figure 2a presents the simulated reflectance spectra of the metasurfaces with varying nanodisk radius $R_2$ under normally incident *x*-polarized plane waves. The excitation of the quasi-BICs manifests as a spectrally narrow resonance with a distinct reflectance peak. In contrast, this peak is absent for genuine BICs, as these mode are completely decoupled from external field and thus cannot be excited by the incident light. The evolution of the reflectance with respect to $R_2$ reveals that the resonance linewidth narrows as $R_2$ approaches $R_1$, indicating a transition from quasi-BICs to genuine BICs. The *Q*-factors are estimated using the eigenfrequencies, $Q = \text{Re}(\omega)/2\text{Im}(\omega)$. As shown in Fig. 2b, the radiative *Q*- factor increases dramatically with $R_2$, corresponding to the recovered translation symmetry. While theoretically unlimited high *Q*-factors can be achieved by tuning $R_2$ closer to $R_1$, practical constraints such as material losses, array size, and fabrication imperfections impose limitations. For demonstration, the quasi-BICs are observed when $R_2$ = 200 nm, and the simulated reflectance spectrum is shown in Fig. 2c. The inset illustrates the electric field distributions at the quasi-BICs, indicating strong field confinement within the metasurfaces, with an enhancement factor over 20, facilitating nonlinear light-matter interactions. Notably, as shown in Fig. 2d, the reflectance spectra remain invariant across polarization angle ranging from 0· to 360·, which essentially comes from the polarization-independent nature of the doubly degenerate quasi-BICs, ensuring robust enhancement regardless of polarization direction.

Building on the numerical design, we conduct experimental demonstrations of the doubly degenerate quasi-BIC metasurfaces. A set of metasurface samples with $R_2$ ranging from 200 nm to 250 nm, in increments of 10 nm, are fabricated using the standard nanofabrication techniques, referring to the Supporting Information for detailed fabrication procedures. The high quality of the fabricated samples is evidenced by the scanning electron microscopy (SEM) images of the samples with $R_2$ = 250 nm and 200 nm, shown in Fig. 3a. The reflectance measurements are performed using a custom-built optical setup, referring to the Supporting Information for detailed description. The measured reflectance spectra of metasurfaces with varying $R_2$ are presented in Fig. 3b, and the extracted *Q*-factors, obtained by fitting the Fano formula[31, 33, 50], are shown in Fig. 3c. It can be observed that as $R_2$ approaches $R_1$, the resonance linewidth of the quasi-BICs narrows under smaller perturbations, resulting in higher *Q*-factorswhich indicate stronger localized electric field enhancement. To confirm the polarization-independent behavior, reflectance spectra are measured for polarization angles ranging from 0· (*x*-polarized incidence) to 90· (*y*- polarized incidence), are shown in Fig. 3d. These measurements confirm the polarization insensitivity of the doubly degenerate quasi-BICs. The experimental results are in good agreement with numerical simulations in terms of spectral positions and polarization independence. However, the observed broader resonance linewidths and the resultant lower *Q*-factors in the experimental spectra, compared to theoretical predictions, are attributed to the losses caused by structural fabrication imperfections and disorder.

Next we proceed to the nonlinear optical responses of the same set of metasurface samples. The THG measurements are conducted in the reflection mode using the optical setup schematically depicted in Fig. 4a. A femtosecond laser system featuring high intensity and ultra-short pulse duration is employed here as the light source to pump the THG process, and two power meters are utilized to measure the powers at the fundamental and THG wavelengths, respectively, referring to the Supporting Information for detailed description. To maximize the resonantly enhanced THG signal, we tune the central wavelength of the pump light to align with the quasi-BIC wavelength of each sample. The normalized conversion efficiency, defined as $\zeta = P_{\text{THG}}^{\text{peak}}/(P_{\text{Pump}}^{\text{peak}})^3$[32, 51], is calculated, where $P_{\text{THG}}^{\text{peak}}$ is the output peak power at the THG wavelength and $P_{\text{Pump}}^{\text{peak}}$ is incident peak power at the pump wavelength. This metric, independent of the pump power, emphasizes the intrinsic role of field enhancement within the metasurfaces in driving nonlinear processes. The results for metasurfaces with varying radius $R_2$ are compared in Fig. 4b, revealing a significant increase in normalized conversion efficiency as $R_2$ approaches R1. This trend is consistent with the localized field enhancement characterized by the *Q* factors in Fig. 3c.

Figure 4c compares the measured THG spectrum of the fabricated metasurface sample with $R_2$ = 240 nm to the unpatterned silicon film of the equivalent thickness, both excited at a pump light central wavelength of 1539 nm matching the metasurface's quasi-BIC wavelength. Both spectrums exhibit a peak at 513 nm, but the THG signal from the quasiBIC metasurface is observed to be approximately 2200 times stronger than that from the silicon film, highlighting the substantial role of highly localized electric fields induced by the quasi-BIC metasurfaces in nonlinear optical effects. We also map the resonantly enhanced THG signals with the polarization directions. In our experiments, the polarization of the pump light is varied by rotating a half-wave plate. The measured results from 0° to 90° are plotted as circles, while the theoretical fittings from 0° to 360° are represented by solid lines in Fig. 4d. As expected, the THG power remains quite stable for any linearly polarized pump, which stems from the polarization-independent nature of the doubly degenerate quasi-BICs. In contrast to previous studies on enhanced harmonic generation via symmetry-protected quasi-BICs, where significant harmonic signals are typically observed only at specific polarization angles due to selection rules for the excitation of the quasi-BICs[52], our results demonstrate a robust enhancement of THG signals that is invariant to the pump polarization. The $C_{4v}$ symmetrical arrangements of the meta-atoms contribute to the polarization-independent optical responses across the linear and nonlinear regimes.

Finally, we examine the power-dependence of the resonantly enhanced THG process. As illustrated in Fig. 4e, the THG power exhibits a significant increase with the pump power. Through fitting the experimental data to the function of $P_{\text{THG}}^{\text{average}} = a(P_{\text{Pump}}^{\text{average}})^b$, we observe a nearly cubic relationship with an exponent $b = 3.08$, which aligns with the anticipated scaling law for THG. The log-log plot further confirms this cubic dependence, with the fitted slope of 3.08 validating the occurrence of the THG process. The pump power ranges from approximately 1 to 3 mW, and the corresponding pump intensity can be estimated considering the pulse duration, repetition rate, and spot size of the fundamental pump light. At an average pump power of 2.91 mW, the average collected THG power reaches up to 29.95 nW. In this case, the conversion efficiency, defined as $\eta = P_{\text{THG}}^{\text{average}}/P_{\text{Pump}}^{\text{average}}$, is found to be $1.03\times10^{-5}$ at a pump intensity of around 5.85 GW/cm². Despite employing a moderate pump intensity, the conversion efficiency is comparable with previously reported experimental results in the literatures[17, 18, 20, 21, 33, 53]. We also notice that the THG power in our measurements does not saturate, and the power dependence of the quasi-BIC metasurfaces closely adheres to the expected cubic relationship.

## III. CONCLUSIONS

In conclusion, we have experimentally demonstrated the robust enhancement of the THG process in silicon metasurfaces empowered by doubly degenerate quasi-BICs. By precisely perturbing the unit cell symmetry of the diatomic metasurfaces, specifically, breaking translation symmetry while preserving $C_{4v}$ rotational symmetry, the quasi-BICs maintain the polarization-independent degenerate behavior in both linear and resonantly enhanced nonlinear optical responses. This results in a notably advantageous polarization-insensitive THG profile, regardless of polarization of the pump light. The maximum THG conversion efficiency achieved is 1.03×10⁻ [5] under a pump intensity of 5.85 GW/cm$^2$, which is comparable to the most state-of-the art performances reported to date. Importantly, the symmetry configuration used in this work is generalizable and can be extended to enhance other nonlinear optical effects at various wavelengths through scalable metasurface designs[54] or integration with two-dimensional materials[55–57] and reconfigurable materials[58–63]. From this perspective, our work offers critical insights into the development of high-efficiency, polarization-independent photon upconversion systems.

**ACKNOWLEDGMENTS**

This work was supported by the National Natural Science Foundation of China ( Grants No. 12304420, No. 12264028, No. 12364045, No. 12364049, and No. 12104105) , the Natural Science Foundation of Jiangxi Province (Grants No. 20232BAB201040, No. 20232BAB211025, and No. 20242BAB25041), and the Young Elite Scientists Sponsorship Program by JXAST (Grants No. 2023QT11 and No. 2025 QT 04).

Tingting Liu and Meibao Qin contributed equally to this work.

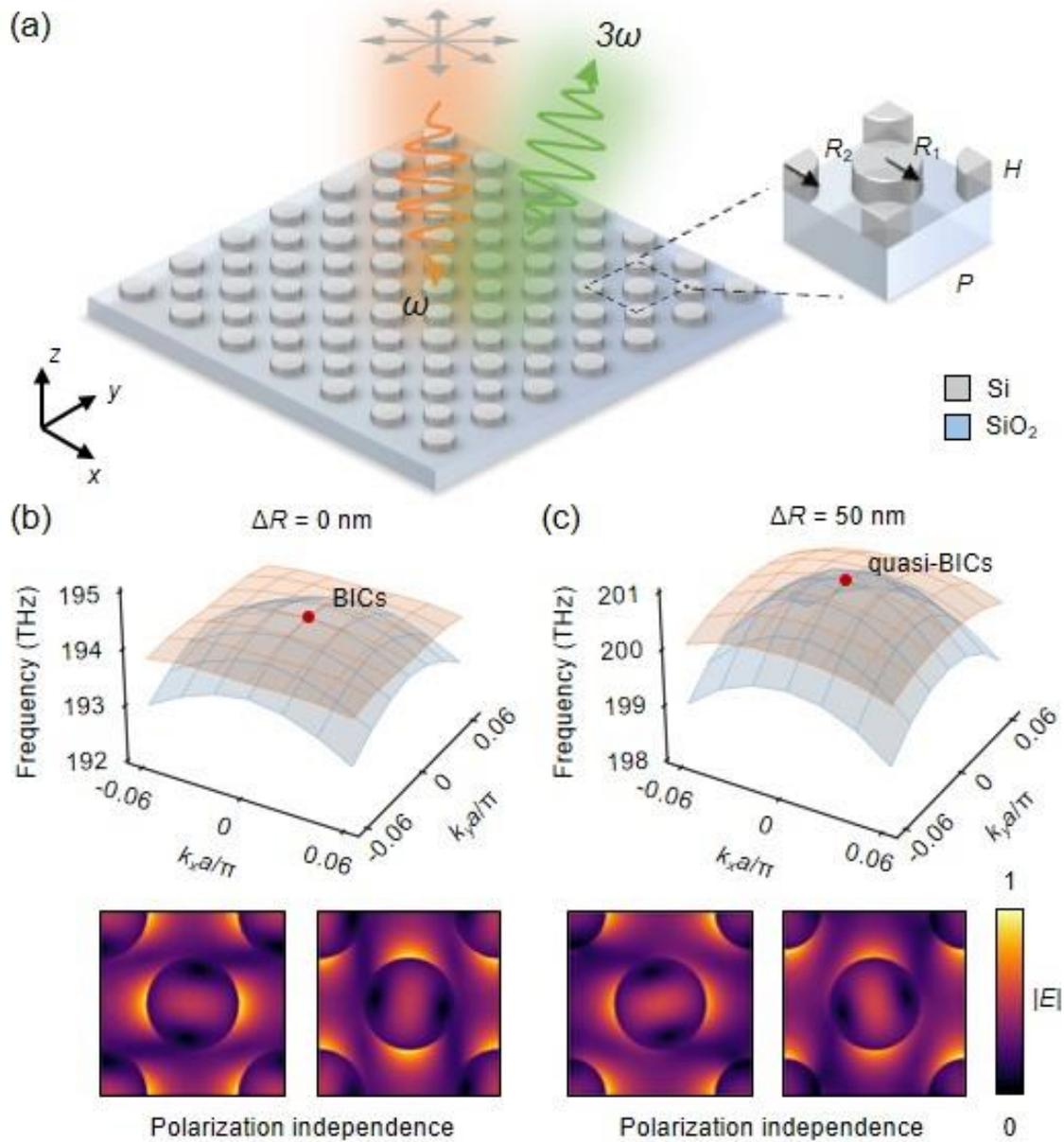

FIG. 1. Enhanced THG in the doubly degenerate quasi-BIC metasurfaces. (a) The schematic of the THG process in the as-designed diatomic metasurfaces consisting of a periodic array of silicon nanodisks placed on top of a glass substrate. (b) The band structure and the electric field distributions in the x-y plane at the BICs with $\Delta R = 0$. (c) The band structure and the electric field distributions in the x-y plane at the quasi-BICs with $\Delta R = 50$ nm.

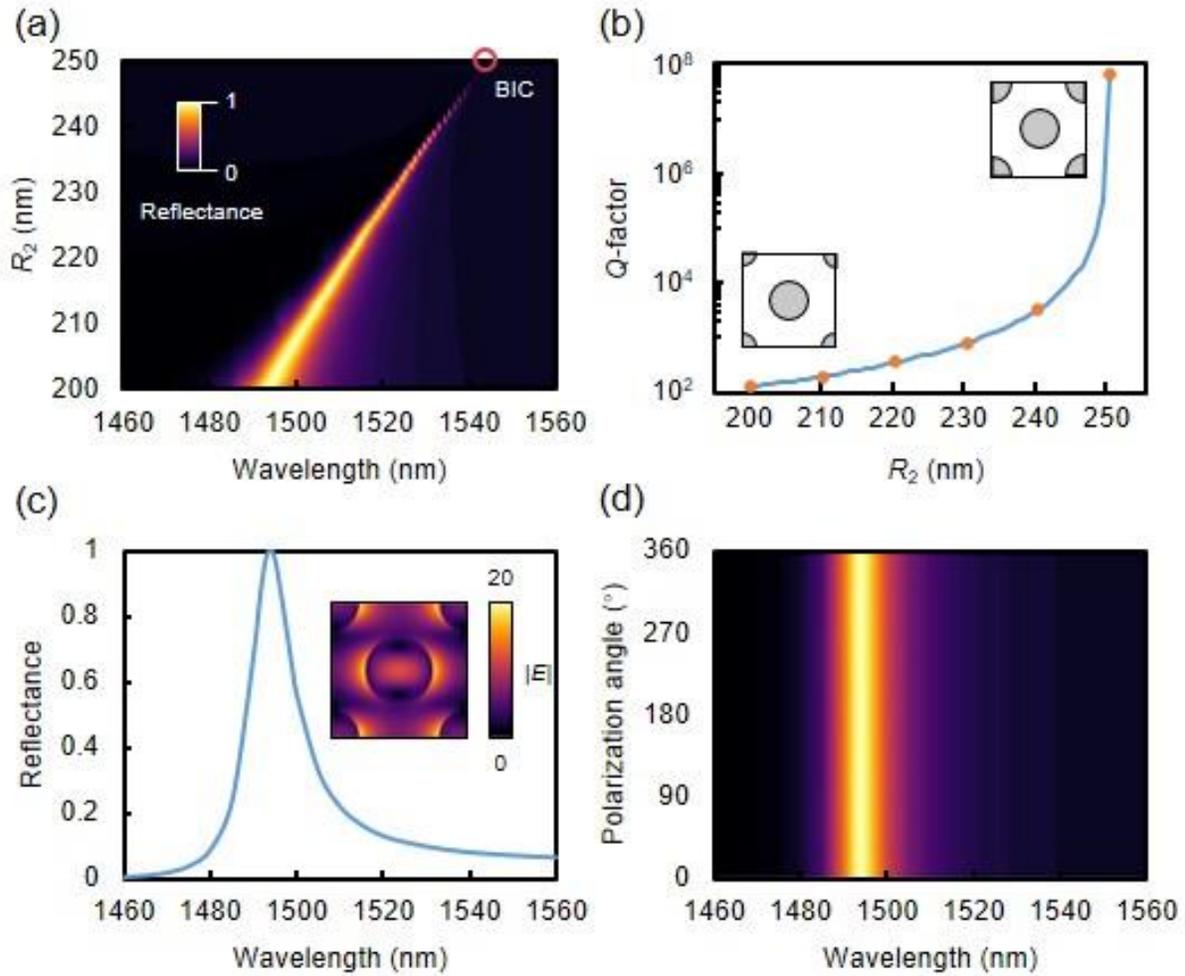

FIG. 2. Simulated linear optical responses of the doubly degenerate quasi-BIC metasurfaces. (a) The reflectance spectra as a function of nanodisk radius $R_2$ under $x$- polarized incidence. The red circle signifies the genuine BICs with $R_2 = R_1 = 250$ nm. (b) The evolution of $Q$-factor of the quasi-BICs with radius $R_2$. (c) The reflectance spectrum and the electric field distribution in the $x$-$y$ plane at the quasi-BICs with radius $R_2$ = 200 nm. (d) The reflectance spectra under different polarization angles with radius $R_2$ =200 nm.

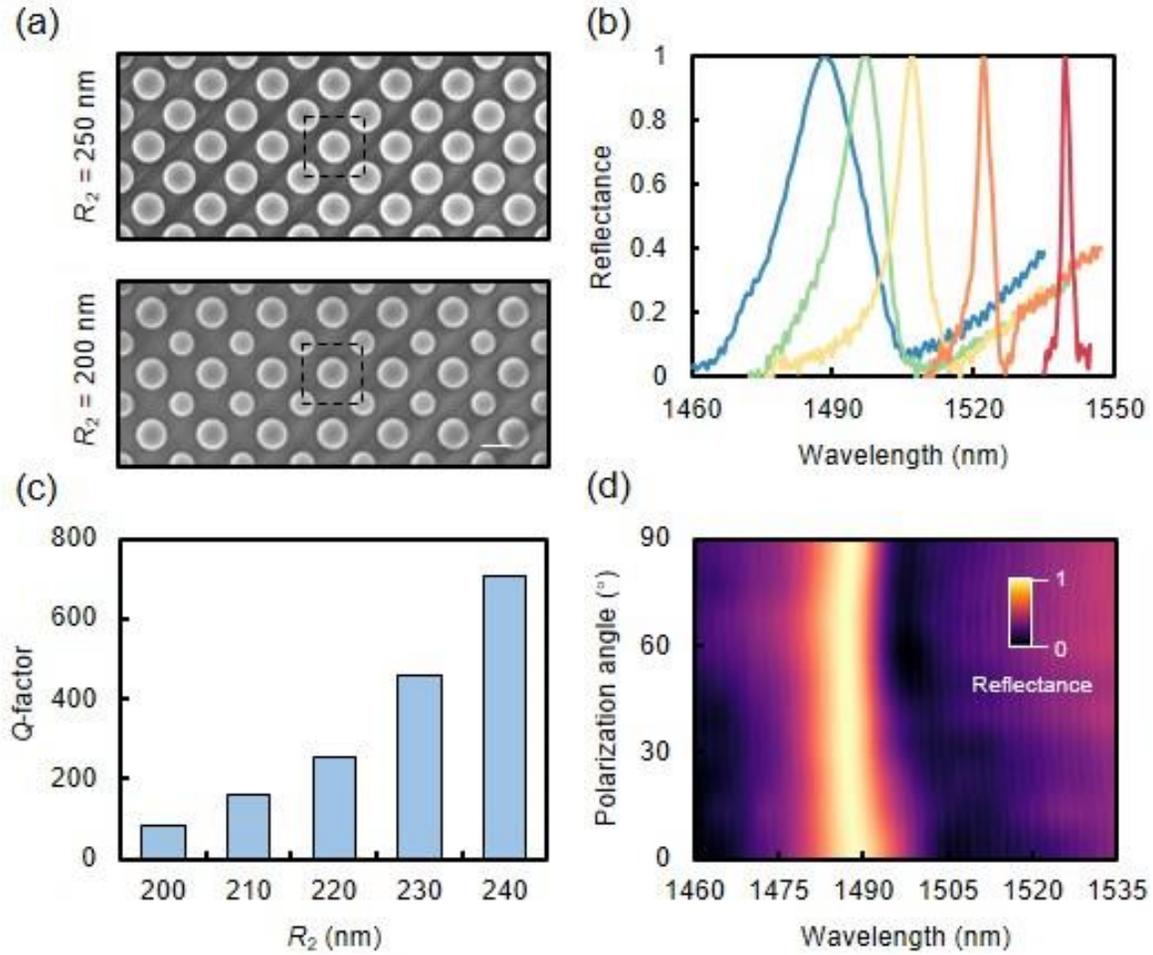

FIG. 3. Experimental measured linear optical responses of the doubly degenerate quasi-BIC metasurfaces. (a) The SEM images of the fabricated metasurface samples in the top view on a scale of 500 nm. (b) The reflectance spectra of the metasurfaces with radius $R_2$ ranging from 200 nm to 240 nm in increments of 10 nm (corresponding to curves from left to right) under $x$- polarized incidence. (c) The $Q$-factors obtained from (b) with radius $R_2$. (d) The reflectance spectra under different polarization angles with radius $R_2$ =200 nm.

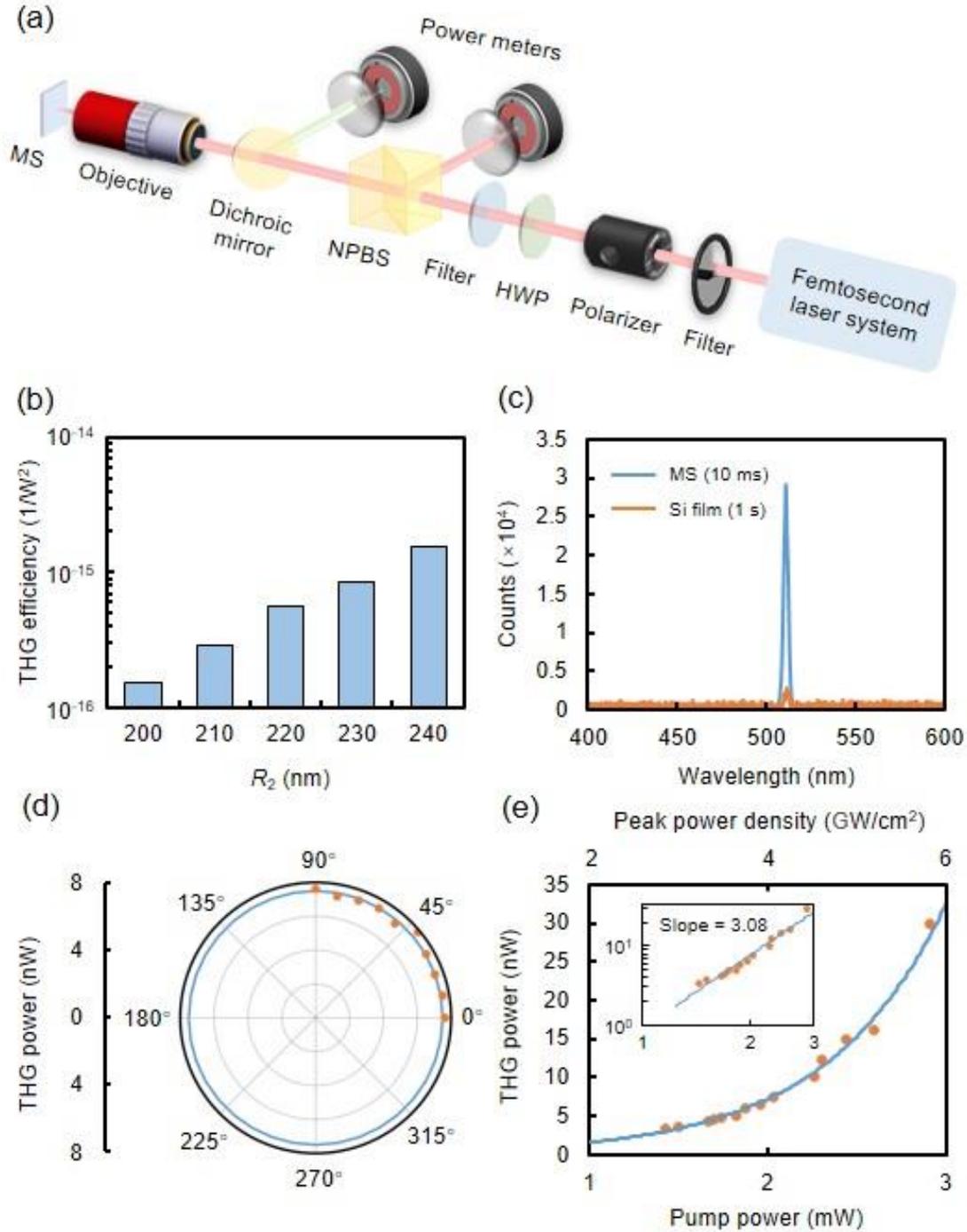

FIG. 4. Experimental measured nonlinear optical responses of the doubly degenerate quasi-BIC metasurfaces. (a) The schematic of the experimental setup for nonlinear THG measurement. (b) The THG normalized conversion efficiency of the metasurfaces with radius $R_2$ ranging from 200 nm to 240 nm under x-polarized incidence. (c) The THG spectra of the metasurfaces with radius $R_2 = 240$ nm and the unpatterned silicon film. (d) The THG power profile under different polarization angles with radius $R_2$

=240 nm. (e) The THG power as a function of pump power with radius R2 =240 nm. Inset shows power dependence of THG process in logarithmic scale.